\documentclass{PoS}

\title{High loop renormalization constants for Wilson
  fermions/Symanzik improved gauge action}

\ShortTitle{High loop renormalization constants for Wilson
  fermions/Symanzik improved gauge action}

\author{\speaker{M.~Brambilla}\\
  Universit\`a di Parma \& INFN, Viale Usberti 7/A, I-43100 Parma, Italy\\
  E-mail: \email{michele.brambilla@fis.unipr.it}}

\author{F.~Di Renzo\\
  Universit\`a di Parma \& INFN, Viale Usberti 7/A, I-43100 Parma, Italy\\
  E-mail: \email{francesco.direnzo@fis.unipr.it}}

\author{L.~Scorzato\\
  ECT*, Strada delle Tabarelle 286, I-38050\ Villazzano~(TN), Italy\\
  E-mail: \email{scorzato@ect.it}}

      \abstract{We present the current status of our computation of quark bilinear renormalization constants for Wilson fermions and Symanzik improved gauge action. Computations are performed in Numerical Stochastic Perturbation Theory. Volumes range from $10^4$ to $32^4$.\\
        Renormalization conditions are those of the RI'-MOM scheme,
        imposed at different values of the physical scale. Having
        measurements available at several momenta, irrelevant effects
        are taken into account by means of hypercubic symmetric
        Taylor expansions. Finite volumes effects are assessed
        repeating the computations at different lattice sizes. In this
        way we can extrapolate our results to the continuum limit, in
        infinite volume.  }

      \FullConference{The XXVII International Symposium on Lattice Field Theory - LAT2009\\
        July 26-31 2009\\
        Peking University, Beijing, China}

\begin{document}

\section{Motivations}
For a long time, lattice perturbation theory was the only available
tool for the computation of Lattice QCD Renormalization Constants
(RC's).  Since the introduction of methods that allow the non
perturbative computation of the RC's for generic composite
operators~\cite{Martinelli:1994ty}\cite{Luscher:1992an}, these
techniques are preferred.  Nevertheless there is no theoretical
obstacle to the perturbative computation of either finite or
logarithmically divergent RC's.  In principle, Lattice Perturbation
Theory (LPT) provides the connection between lattice simulations and
continuum perturbative QCD, that works only at high energy.  The main
difficulties of LPT are actually practical.  First of all, LPT
requires much more effort than in the continuum. Because of this, a
lot of perturbative results are known only at one loop. Second, LPT
series present bad convergence proprieties, so that we should
emphasize that \emph{a fortiori} one loop computations cannot give the
correct results.  A good example is provided by the computation of the
quark mass renormalization constant: there appear to be discrepancies
in between determinations coming from perturbative and
non-perturbative techniques~\cite{Blossier:2007vv}.  The very point is
that as long as a comparison is made taking into account one loop, it
is virtually impossible to assess the
systematics of both results.\\
In order to obtain higher orders in PT expansions, one can make use of
Numerical Stochastic Perturbation Theory (NSPT)~\cite{Di
  Renzo:2004ge}, a numerical implementation of Stochastic
PT~\cite{Parisi:1980ys}. Renormalization constants can be computed in
NSPT at 3 (or even 4) loops, as it has been done in the case of Wilson
fermions with non-improved gauge actions~\cite{Di Renzo:2006wd}. A
nice aspect is that one can work in the massless limit, where RC's are
often defined, and no chiral extrapolation is needed. We will discuss
how one can assess to finite $a$ effects by means of \emph{hypercubic
  Taylor expansion}. Moreover, finite volume effects can be corrected
by repeating the computations at different lattice sizes. Measuring
RC's for different values of $n_f$ is also possible to know the
dependence on the number of flavors.

\section{RI'-MOM scheme}
The scheme we will adhere to is the so called RI'-MOM scheme, which
became much popular since the development of non-perturbative
renormalization. RI emphasizes that the scheme is \emph{regulator
  independent}, which makes the lattice a viable regulator, while the
prime signals a choice of renormalization conditions which is slightly
different from the original one. An important feature of this scheme
is the fact that the relevant
anomalous dimensions are known up to three loops in the literature~\cite{Gracey:2003yr}.\\
The main observables in our computations are the quark bilinears
between states at fixed off shell momentum $p$:
\begin{equation}\label{Propagator}
\int dx\langle p|\bar\psi(x)\Gamma\psi(x)|p\rangle = G_\Gamma(pa)
\end{equation}
where $\Gamma$ stands for any of the 16 Dirac matrices, returning 
the S,~V,~P,~A,~T currents. Our notation points out the $pa$ lattice space
dependence.\\
Since these quantities are gauge dependent we need to fix the
gauge. We will work in Landau gauge, which is easy to fix on the
lattice~\cite{Davies:1987vs}. This gauge condition also gives an
advantage: the anomalous dimension for the quark field is zero at one
loop.\\
Given the quark propagator $S(pa)$, one can obtain the amputated
functions
\begin{equation}\label{Amputated}
G_\Gamma(pa)\to\Gamma_\Gamma(pa) = S^{-1}(pa)G_\Gamma(pa)S^{-1}(pa).
\end{equation}
The $\Gamma_\Gamma(pa)$ are then projected on the tree-level structure
by means of a suitable projector $\hat P_{O_\Gamma}$
\[
O_\Gamma(pa) = \mbox{Tr}\left( \hat P_{O_\Gamma}
  \Gamma_\Gamma(pa)\right).
\]
We can finally express the renormalization conditions in terms of the
$O_\Gamma(pa)$ operator:
\[
\left.Z_{O_\Gamma}(\mu a,g(a))Z^{-1}_q(\mu
  a,g(a))O_\Gamma(pa)\right|_{p^2=\mu^2} = 1.
\]
The $Z$'s depend on the scale $\mu$ via the dimensionless parameter
$\mu a$, while the dependence on the coupling $g(a)$ is given by the 
perturbative expansion. The quark field renormalization constant
$Z_q$ in the formula above is defined by
\begin{equation}\label{Zq:def}
\left.Z_q =
  -i\frac{1}{12}\frac{\mbox{Tr}(\slash\!\!\!\!\!pS^{-1}(pa))}{p^2}\right|_{p^2=\mu^2}.
\end{equation}
To obtain a mass independent scheme, we impose these conditions at the
massless point. In the case of Wilson fermions this requires the knowledge of
the critical mass. While one and two loop results are known
from the literature (and can be used as a consistency test) the third loop 
is a byproduct of these computations.

\section{Finite lattice size effects}
At the generic n$th$-loop the RC's take the form
\[
z_n = c_n + \sum_{j=1}^n d_j(\gamma)\log^j(pa) + F(pa),
\]
where logarithmic contributions are known from the literature and one is 
mainly interested in the finite number $c_n$. The first thing
to do is then to subtract the divergent $\log$s (again, 
we take them from the literature). After such a
subtraction we still have the irrelevant contribution $F(pa)$, that
can be fitted by means of a hypercubic Taylor expansion. We show how
this technique works by an example.\\
Consider the two points function in
the continuum limit:
\[
\Gamma_2(p^2) = S^{-1}(p^2).
\]
On the lattice it depends on the dimensionless quantity
$\hat{p} = pa$. Furthermore, we explicit write the dependence on the coupling
\begin{eqnarray*}
  \hat{\Gamma}_2(\hat{p}, \hat{m}_{cr},\beta^{-1}) & = & \hat{S}^{-1}(\hat p, \hat{m}_{cr},\beta^{-1}) \\
  & = & i\hat{\slash\!\!\!\!\!p} + \hat{m_W}(\hat p) - \hat{\Sigma}(\hat{p}, \hat{m}_{cr},\beta^{-1}),
\end{eqnarray*}
where the $hat$ stands for a dimensionless quantity. Here $\hat m_W$
is the mass term generated by the Wilson prescription (it is $\mathcal
O(\hat{p}^2)$), $\hat{\Sigma}(\hat{p}, \hat{m}_{cr},\beta^{-1})$ is the self
energy and $\hat m_{cr}$ is the quark critical mass. The fermion mass
counterterm arises because Wilson regularization breaks chiral
symmetry. Both these last two terms are $\mathcal O(\beta^{-1})$.\\
The self energy can be written as
\[ \hat{\Sigma}(\hat{p}, \hat{m}_{cr},\beta^{-1}) =
\hat{\Sigma}_c(\hat{p}, \hat{m}_{cr},\beta^{-1}) +
\hat{\Sigma}_V(\hat{p}, \hat{m}_{cr},\beta^{-1}) +
\hat{\Sigma}_{other}(\hat{p}, \hat{m}_{cr},\beta^{-1}).
\] These are the different contribution along the different elements
of the Dirac base. In particular, $\hat{\Sigma}_c$ is the contribution
along the identity, $\hat{\Sigma}_V$ the contribution along the gamma
matrices.\\ The self energy $\hat{\Sigma}_c$ contains the contribution of
the mass:
\[ \hat{\Sigma}(0, \hat{m}_{cr},\beta^{-1}) = \hat{\Sigma}_c(0,
\hat{m}_{cr},\beta^{-1}) = \hat m_{cr} = am_{cr}.
\] To show how hypercubic Taylor expansions work, we can concentrate
on the $\hat\Sigma_V$ term, from which we can in this way extract the
quark field RC. We can consider a Taylor expansion of $\hat\Sigma_V$
in the form
\begin{equation}\label{SigmaV:expansion}
\hat\Sigma_V(\hat{p}, \hat{m}_{cr},\beta^{-1}) =
i\sum_\mu\gamma_\mu\hat p_\mu\left( \hat\Sigma_V^{(0)}(\hat{p},
\hat{m}_{cr},\beta^{-1})+ \hat p^2_\mu\hat\Sigma_V^{(1)}(\hat{p},
\hat{m}_{cr},\beta^{-1})+ \hat p^4_\mu\hat\Sigma_V^{(2)}(\hat{p},
\hat{m}_{cr},\beta^{-1})+ \ldots \right)
\end{equation}
where the expansion entails an expansion in powers of $a$. The
functions $\hat\Sigma_V^{(i)}(\hat{p}, \hat{m}_{cr},\beta^{-1})$ are
combinations of hypercubic invariants. As an example, the first term
in the expansion~(\ref{SigmaV:expansion}) can be written as
\begin{equation}\label{SigmaV0:expansion}
\hat\Sigma_V^{(0)}(\hat{p}, \hat{m}_{cr},\beta^{-1}) = 
\alpha_1^{(0)} 1 +
\alpha_2^{(0)} \sum_\nu p_\nu^2 +
\alpha_3^{(0)} \sum_\nu p_\nu^4 + \alpha_4^{(0)} \left(\sum_{\nu} p_\nu^2\right)^2 +
\mathcal{O}(a^6)
\end{equation}
As a general recipe, all the possible covariant polynomials can be
found via a character's projection of the polynomial representation of
the hypercubic group onto the defining representation of the group. 
One can see that plugging this expansion in the definition~(\ref{Zq:def}) of
$Z_q$ the only term that doesn't vanish in the continuum limit is
$\alpha_1^{(0)}$, the first coefficient of the expansion of
$\hat\Sigma_V^{(0)}$.

\section{Finite volume effects}
In the previous section we have shown how one can overcome the effects
due to the discretization. These are not the only systematic effects in a
lattice simulation. Though large lattices are available, one
can't neglect the finite volume effects. This means that on a finite
lattice we have to consider also a $pL$ dependence in our
quantities.\\
We will now show how the analysis of $\Sigma = \hat{\Sigma}(pa, pL,
am_{cr},\beta^{-1})$ can be modified to take care of the finite volume
effects. Let us take $\hat{\Sigma}^{(0)}_V(pa, pL,
am_{cr},\beta^{-1})$ as an example. In the spirit of
  \cite{Kawai:1980ja}, consider the ansatz
\begin{eqnarray*}
  \hat{\Sigma}^{(0)}_V(pa, pL) & = & \hat{\Sigma}^{(0)}_V(pa) + \left(\hat{\Sigma}^{(0)}_V(pa, pL) - \hat{\Sigma}^{(0)}_V(pa) \right)\\
  & = & \hat{\Sigma}^{(0)}_V(pa) + \Delta\hat{\Sigma}^{(0)}_V(pa, pL)
\end{eqnarray*}
where $\hat{\Sigma}^{(0)}_V(pa) = \hat{\Sigma}^{(0)}_V(pa, \infty)$
and to keep the formula simpler we omit the dependence on 
$\hat m_{cr}$ and $\beta^{-1}$ .\\
We can now expand as in~(\ref{SigmaV0:expansion}):
\begin{equation}\label{SigmaVpL}
\hat\Sigma_V^{(0)}(\hat{p}, pL) = 
\alpha_1^{(0)} 1 +
\alpha_2^{(0)} \sum_\nu p_\nu^2 +
\alpha_3^{(0)} \sum_\nu p_\nu^4 + \alpha_4^{(0)} \left(\sum_{\nu} p_\nu^2\right)^2 +
\Delta\hat{\Sigma}^{(0)}_V(pL) +
\mathcal{O}(a^6)
\end{equation}
where the continuum limit $\Delta\hat{\Sigma}^{(0)}_V(pL) =
\Delta\hat{\Sigma}^{(0)}_V(pa = 0, pL)$ has been taken into
account. The rationale for this is the following: $pL$ effects are
present also in the continuum limit, and $pa$ corrections on top
of those are regarded as \emph{corrections on corrections}. In principle, also
these corrections can be taken into account, but this requires a
larger number of parameters.\\
The latter assumption has a strong consequence: measurements on
lattices of different size are affected by the same $pL$ effect once
one consider the same tuples $\left(n_1,n_2,n_3,n_4\right)$
\[
p_\mu L = \frac{2 \pi n_\mu}{L} L = 2\pi n_\mu.
\]
The practical implementation is the following. First of all one has to
select a collection of lattice sizes and an interval
$[(pa)^2_{min},(pa)^2_{max}]$. Then one considers all the tuples $\vec
n = \left(n_1, n_2, n_3, n_4\right)$ which, for all the lattice sizes,
fall in $\left[(pa)^2_{min},(pa)^2_{max}\right]$ and chooses one
representative for all the tuples connected by an H4
transformation. We further add to these data the measure taken at the
highest value of $(pa)^2$ which falls in the interval
$\left[(pa)^2_{min},(pa)^2_{max}\right]$ on the lattice with the
biggest value of $N=L/a$. Assuming that this tuple
(which we will call $n^*$) is a good approximation to $pL = \infty$, the measure will be considered as a normalization point. Is then possible to fit the parameters in the expansion (in our case the $\alpha_i^{(0)}$), and the finite volume corrections, which came in the same number of the considered tuples.\\
To be explicit,~(\ref{SigmaVpL}) can then be formulated in terms of
the $(\vec n, N)$ dependence:
\begin{equation}\label{SigmaVn}
\hat\Sigma_V^{(0)}(\vec n\neq\vec{n^*}, N) = 
\alpha_1^{(0)} 1 +
\alpha_2^{(0)} p^2(\vec n, N) +
\alpha_3^{(0)} p^4(\vec n, N) + \alpha_4^{(0)} \left(p^2(\vec n, N)\right)^2 +
\Delta^{(0)}_{\vec n} +
\mathcal{O}(a^6)
\end{equation}
where $\vec n^*$ means the tuple taken as normalization point. For the measure taken at the tuple $\vec n^*$ the formula is slightly different:
\begin{equation}\label{SigmaVnstar}
\hat\Sigma_V^{(0)}(\vec{n^*}, N) = 
\alpha_1^{(0)} 1 +
\alpha_2^{(0)} p^2(\vec{n^*}, N) +
\alpha_3^{(0)} p^4(\vec{n^*}, N) + \alpha_4^{(0)} \left(p^2(\vec{n^*}, N)\right)^2 +
\mathcal{O}(a^6)
\end{equation}
In~(\ref{SigmaVn}) and~(\ref{SigmaVnstar}) $p^2 = \sum_\nu p_\nu^2$ and $p^4 = \sum_\nu p_\nu^4$.
Two issues of the procedure can only be assessed a posteriori: the
assumption that the measure at the normalization point is free from
finite volume effects and the stability of the fit.\\

\section{Results}
\begin{figure}[!htbc]
\begin{centering}
\includegraphics[width=0.7\textwidth]{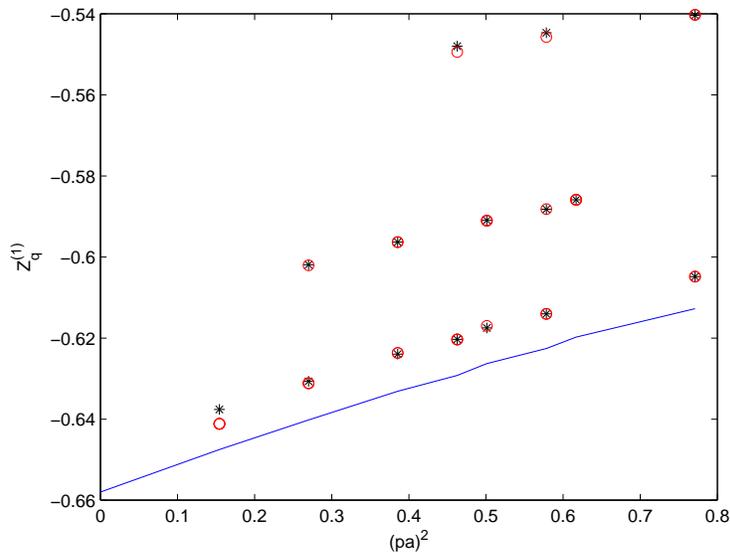}
\end{centering}
\caption{Computation at one loop of $Z_q$. Different ``families'' of
  point correspond to the different components of momentum $p$ along
  the relevant direction. In red measured values at given
  $(p^2,p_\mu)$, in red the fit results and in blue the extrapolation
  at $a = 0$.}
\label{Zq}
\end{figure}

\begin{figure}[!htbc]
\begin{centering}
\includegraphics[width=0.7\textwidth]{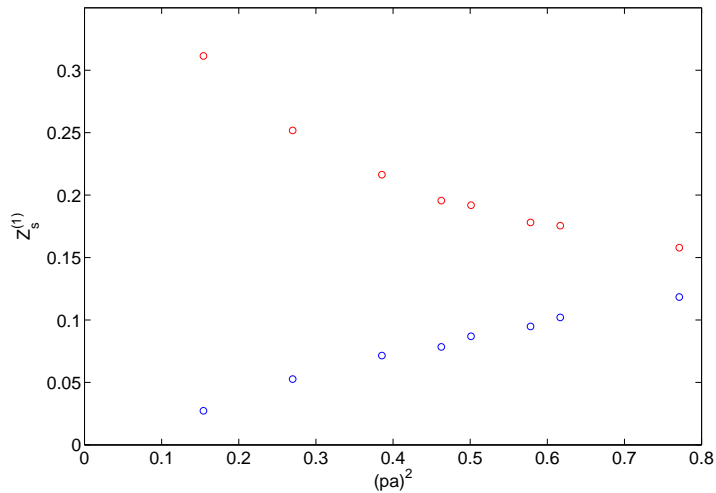}
\end{centering}
\caption{Computation at one loop of $Z_s$. The anomalous dimension at
  one loop is different from zero, and the measure shows a logarithmic
  divergence (red points). In blue the results after the
  $\gamma_s\log{\hat p}^2$ contribution subtraction.}
\label{Zs}
\end{figure}

\begin{figure}[!htbc]
\begin{centering}
\includegraphics[width=0.7\textwidth]{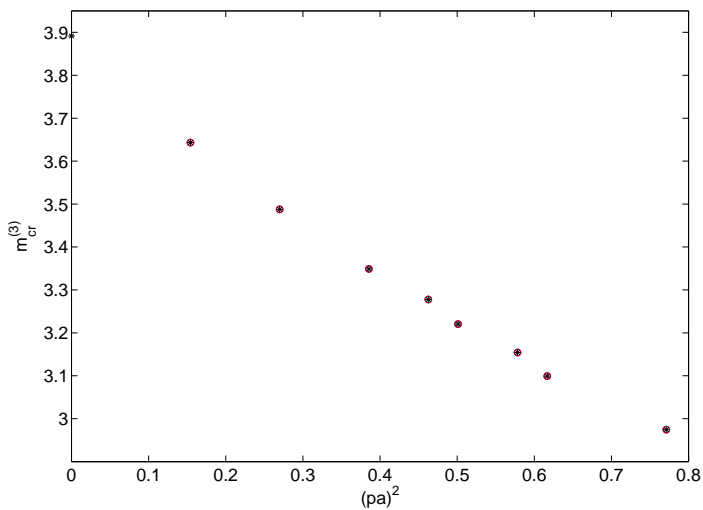}
\end{centering}
\caption{Preliminar result for the critical mass at three loop. In red the result of measurements, in black the result of the fitting procedure.}
\label{mcr}
\end{figure}

In the previous sections, we discussed the procedure to obtain results and keep finite $a$ and $pL$ effects under control. Since the work is still in progress, we only display the first technique at work. We computed at every order relevant expectation values given by~(\ref{Propagator}) and amputated to obtain~(\ref{Amputated}). Finally we projected on the tree level structure and performed the $pa$ analysis by means of the hypercubic Taylor expansion.\\
We are not presenting any new result (apart a preliminar value for
critical mass at three loop), but we can compare some numerical results at leading order with analytical results~\cite{Aoki:1998ar}.\\
The first quantity we consider is the simpler vector one: the measure
of the RC for the quark propagator. This is a log free quantity, so we
only need to extract the correct $a\to 0$ limit. In figure~\ref{Zq} we
show the computation of $Z_q$. It's easy to recognize the effect of
different lengths of $\hat p_\mu$ in the relevant direction.\\
As an example of a logarithmically divergent quantity, in
figure~\ref{Zs} we show the computation of $Z_s$. In the figure, red
point are the measurements before the $\log$ subtraction, blue the
result after the $\gamma_s^{(1)}\log{\hat{p}^2}$ subtraction. Since
this is a scalar quantity, we haven't the effect of the length of the
components in a given direction.\\
In figure~\ref{mcr} we present the preliminar result for the critical
mass at three loop, $m_{cr}^{(3)}$. This quantity doesn't present
either vector structure or $\log$ divergences. Critical mass is a
byproduct of all the previous computations. The introduction of this
counterterm is needed in the computation of the next loop quantities.

\section{Work in progress}
The aim of this work is to compute high order renormalization
coefficients for masses, fields and bilinears for different
actions. Analytic results are known at most up to 2 loops, while there
are NSPT results for Wilson gauge-Wilson fermions up to 3 loops at
various $n_f$. We are in the process of taking into account $pL$
effects for the latter action. We are now applying the method also to
Tree Level Symanzik (current work) and Iwasaki gauge actions. For
these actions we're also interested in computing the $n_f$
dependence. Current data are obtained from $32^4$ lattices measured on
an APE machine, while a C++ code is working on smaller lattices on
standard workstations.

\end{document}